\documentclass[aps,prl,twocolumn,showpacs,superscriptaddress]{revtex4}
\usepackage[pdftex]{graphicx}
\usepackage{amssymb}
\begin{document}

\title{Curvature Fields, Topology, and the Dynamics of Spatiotemporal Chaos}

\author{Nicholas T.~Ouellette}
\affiliation{Department of Physics, Haverford College, Haverford, PA 19041}
\author{J.~P.~Gollub}
\email[Email: ]{jgollub@haverford.edu}
\affiliation{Department of Physics, Haverford College, Haverford, PA 19041}
\affiliation{Department of Physics, University of Pennsylvania, Philadelphia, PA 19104}

\date{\today}

\begin{abstract} 
The curvature field is measured from tracer particle trajectories in a
two-dimensional fluid flow that exhibits spatiotemporal chaos, and is used to
extract the hyperbolic and elliptic points of the flow.  These special points
are pinned to the forcing when the driving is weak, but wander over the domain
and interact in pairs at stronger driving, changing the local topology of
the flow.  Their behavior reveals a two-stage transition to spatiotemporal
chaos: a gradual loss of spatial and temporal order followed by an abrupt onset
of topological changes.  
\end{abstract}


\maketitle

When a system governed by nonlinear equations of motion is driven out of
equilibrium, a variety of complex behaviors can result, ranging from chaos in
low-dimensional systems \cite{Ott1981} to the seemingly random dynamics of
turbulent flows \cite{Falkovich2001}. If the number of active degrees of
freedom (roughly corresponding to the number of equations required to
characterize the dynamics) is small, the mathematics of dynamical systems has
proved to be a powerful tool. When it is quasi-infinite and the dynamics are
turbulent, statistical approaches based on assumed scale invariance have been
fruitful. For spatially-extended systems below the turbulence transition,
however, a regime with a large but finite number of active degrees of freedom
exists that is disordered both in space and in time
\cite{Ciliberto1988,Rabaud1990,Shraiman1992,Cross1993,Dennin1996,Gollub1999,Egolf2000}.
This regime of spatiotemporal chaos (STC) remains poorly characterized and
understood; indeed, there is no generally agreed-upon quantitative indicator of
STC.

In this Letter, we study the dynamics of a simple fluid flow that exhibits STC
by considering its underlying topology. We describe a method for locating the
time-dependent topologically special points of the flow, and show that their
dynamics describe the flow pattern as a whole. These special points undergo
pairwise interactions, changing the flow topology when they are created or
annihilated. Surprisingly, we find that the rate of creation or annihilation
shows a discrete onset, while other measures of spatial and temporal disorder
increase smoothly with the rate of energy input.  This form of STC is
characterized not only by disorder but also by constantly changing flow
topology.

In a driven body of fluid, there can exist instantaneous stagnation points
where the velocity vanishes, relative to some observer. These special points
carry the bulk of the information contained in the flow field: if the locations
of all of these topologically special points and their local flow properties
are known, most of the full flow field can be determined \cite{Perry1987}.
They are also an essential component of chaotic mixing \cite{Jana1992}.  These
points are distinct from the topological defects previously considered in
studies of STC \cite{Egolf1998,Daniels2002,Young2003}, as they are present even
when the flow is not chaotic. In a two-dimensional (2D) flow field, the special
points come in two types. When embedded in a region of the flow that is
dominated by vorticity, they are elliptic; in a strain-dominated region, they
are hyperbolic (\textit{i.e.}~saddle-like). These special points, however, have
proved to be very difficult to identify, particularly in experimental flows. In
this Letter, we show that by considering the curvature of Lagrangian
trajectories, that is, the trajectories of individual moving fluid elements, we
can find the elliptic and hyperbolic points in an automated way, even when they
move. Once located, the trajectories and statistics of the special points give
insight into the transition to and dynamics of STC.

\begin{figure*}
\begin{center}
\includegraphics[width=6in]{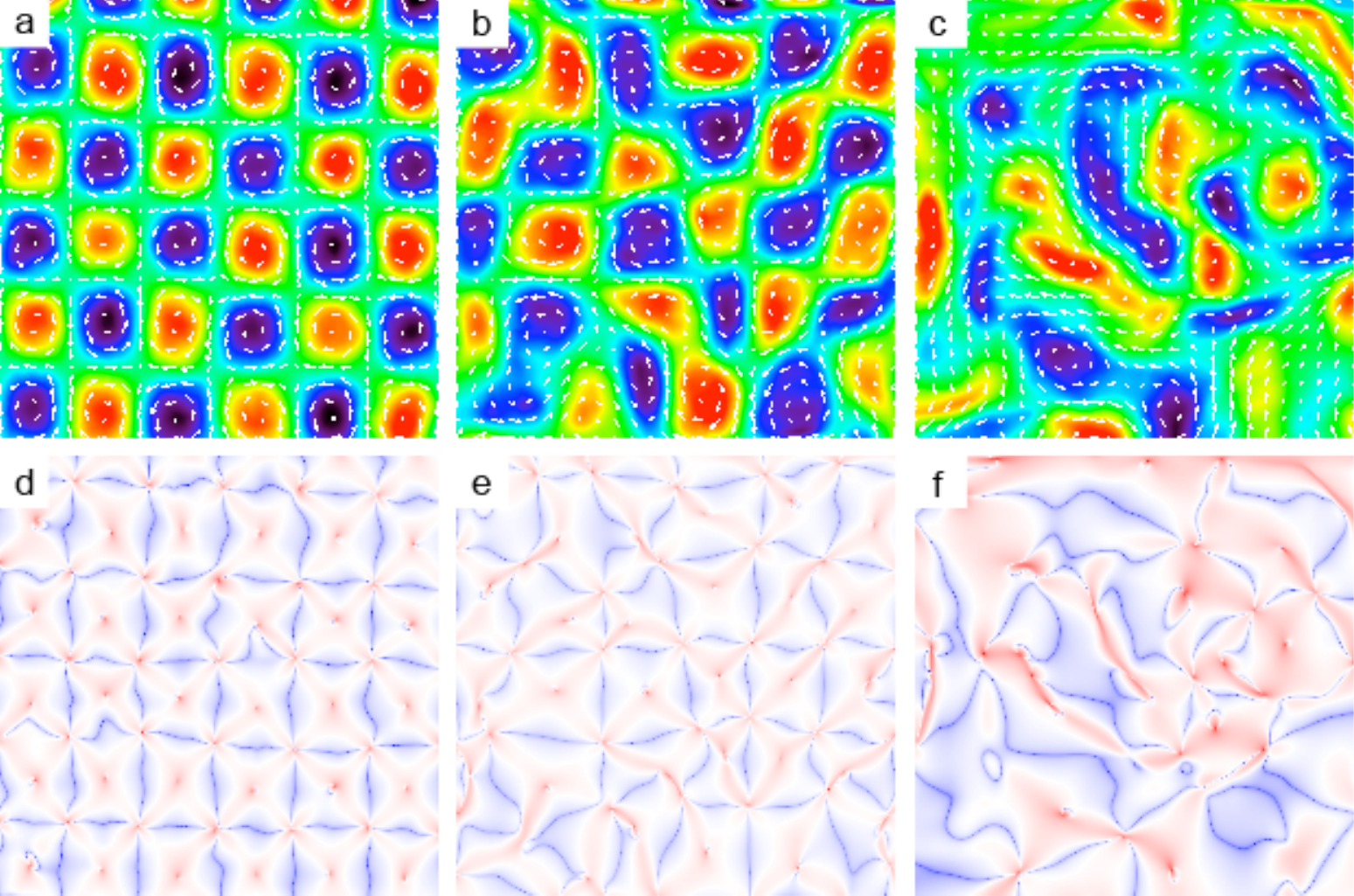}
\caption{\label{fig:fields} (color online) Velocity, vorticity, and curvature fields for $\mathrm{Re} = 32$ (a, d), 93 (b, e), and 245 (c, f). In (a-c), velocity vectors are shown as arrows, undersampled by a factor of 8 for clarity. The vorticity is shown by color: red (or gray) corresponds to large negative vorticity (clockwise rotation), and blue (or black) to large positive vorticity (counterclockwise rotation). As $\mathrm{Re}$ increases, the flow becomes more disordered. (d-f) show the logarithm of the curvature; red corresponds to large and blue to small values. Low values of curvature typically form lines, while the highest values appear as points.}
\end{center}
\end{figure*}

We generate a quasi-2D flow using magnetohydrodynamic forcing in a thin layer
of conducting fluid, as described previously \cite{Rothstein1999,Paret1997}. A
6 mm layer of water containing 8\% by weight of CuSO$_4$ was placed above a
square lattice of permanent magnets with alternating orientation. When a
current is driven across the cell, Lorentz forces set the fluid into motion. The
dimensionless strength of the forcing is measured by the Reynolds number
$\mathrm{Re} = UL/\nu$, where $U$ is the root-mean-square velocity, $L = 2$ cm
is the mean magnet spacing, and $\nu$ is the kinematic viscosity. At low
$\mathrm{Re}$, the flow is a square array of vortices of alternating sign, as
shown in Fig.~1. As $\mathrm{Re}$ grows, however, the flow deviates from the
forced lattice and becomes spatiotemporally chaotic. To measure the flow, we
follow the simultaneous trajectories of thousands of neutrally-buoyant 116
$\mu$m fluorescent polystyrene tracer particles, using algorithms similar to
those described by Ouellette \textit{et al.}~\cite{Ouellette2006}. The
particles are imaged at a rate of 12 Hz, and their positions are determined to
a precision of 25 $\mu$m (0.1 pixels). The velocities and accelerations of the
particles are then computed by fitting polynomials to short segments of the
trajectories \cite{Voth2002}. Statistics are collected in a 10 cm $\times$ 10
cm window in the center of the flow, so that boundary effects are excluded.

To find the elliptic and hyperbolic points, we first compute the curvature
along the trajectories of the tracer particles. Curvature, a geometrical
quantity containing, in principle, no dynamical information, completely
specifies a curve in 2D space. Because of the nature of our measurement
technique, however, the  trajectories are parameterized by time. In this case,
the Frenet formulas show that the curvature is given by  $\kappa = a_n / u^2$,
where $a_n$ is the acceleration normal to the direction of motion and $u$ is
the velocity of the particle. The single-point statistics of curvature have
previously been studied in turbulent flow \cite{Braun2006,Xu2007}, but were
found to be explainable with a simple model that should also apply to
non-turbulent flows. In our flow, we measure curvature probability density
functions (PDFs) that are consistent with the model proposed by Xu \textit{et
al.}~\cite{Xu2007}.

Instead of focusing on such single-point curvature statistics, we consider
curvature fields, analogous to velocity or vorticity fields. Sample curvature
fields for the steady and STC regimes are shown in Fig.~1, and striking
structure is evident. As shown previously in studies of turbulence, the
distribution of curvature is exceptionally wide \cite{Braun2006,Xu2007}.  What
was not observed in previous studies, however, is the tendency of low values of
curvature to be spatially organized into lines, while high values exist as
solitary points. Comparing the vorticity and curvature fields in Fig.~1, we see
that these high-curvature points correspond to the hyperbolic and elliptic
points of the flow. This observation has a clear physical interpretation: near
both hyperbolic and elliptic points, the direction of fluid particle
trajectories changes over very short length scales, corresponding to intense
curvature. By locating the local maxima (with values larger than the mean) of
the curvature field, we can therefore find the topologically special points of
the flow. To classify them as elliptic or hyperbolic, we make use of the
Okubo-Weiss parameter $\Lambda = (\omega^2 - s^2) / 4$, where $\omega^2$ is the
enstrophy and $s^2$ is the square of the strain rate
\cite{Babiano2000,Rivera2001}. If a curvature maximum lies in a region with
$\Lambda > 0$, where rotation dominates the flow, we classify it as elliptic;
if $\Lambda < 0$, the local flow is dominated by strain and we classify the
point as hyperbolic.

\begin{figure}
\begin{center}
\includegraphics[width=3.2in]{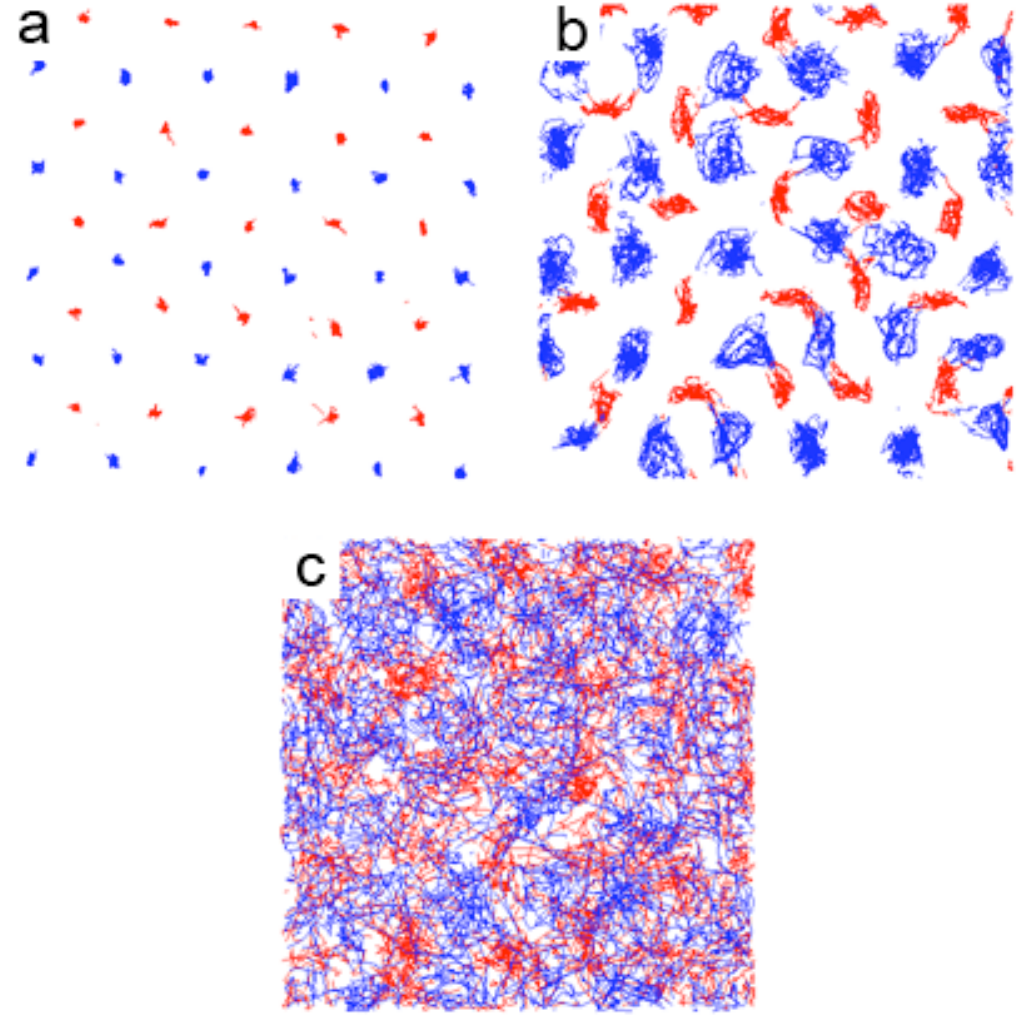}
\caption{\label{fig:tracks} (color online) Trajectories of the topologically special points for the three Reynolds numbers in Fig.~1. Hyperbolic points are plotted in red, while elliptic points are in blue. At $\mathrm{Re} = 32$ (a), the special points are tightly bound to the forced vortex lattice. At $\mathrm{Re} = 93$ (b), they remain bound but make larger excursions. At $\mathrm{Re} = 245$ (c), where the flow is spatiotemporally chaotic, the special points wander over the domain.}
\end{center}
\end{figure}

Once we have located the hyperbolic and elliptic points at each instant in
time, we can feed their positions into the same tracking algorithms we use to
construct the tracer-particle trajectories, and thereby study their dynamics.
At low $\mathrm{Re}$, where the underlying flow is a vortex array locked to the
magnetic forcing, the hyperbolic and elliptic points lie on a square lattice,
as shown in Fig.~2, with the elliptic points in the vortex centers and the
hyperbolic points at the vortex corners. As $\mathrm{Re}$ is increased, the
special points wander progressively farther from their lattice sites. We
observe that the elliptic points move in quasi-circular orbits, while the
hyperbolic points move primarily along their stable manifolds. Finally, when
$\mathrm{Re}$ is high enough, the special points break free from the lattice
and wander freely; in a sense, the lattice melts. We note that the dynamics of
the hyperbolic and elliptic points are representative of the dynamics of the
flow pattern, rather than of the tracer particles themselves; the individual
tracer particles may wander between the vortex cells even in the regime where
the special points remain pinned.

\begin{figure}
\begin{center}
\includegraphics[width=3.2in]{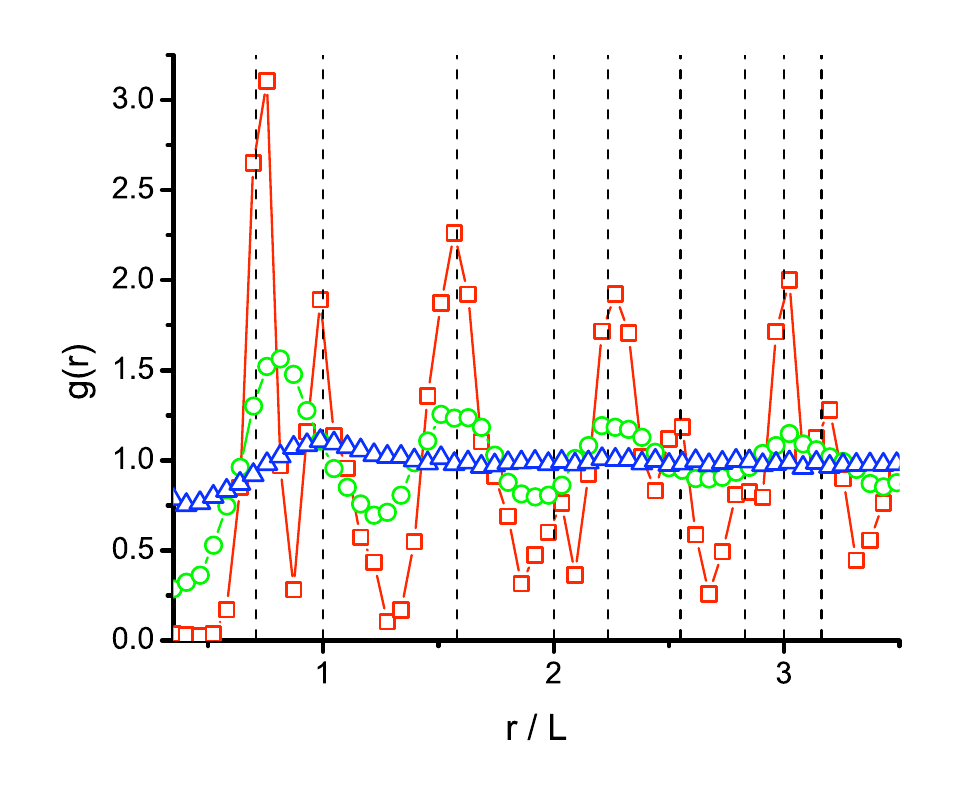}
\caption{\label{fig:separation} (color online) The radial distribution function $g(r)$ of the topologically special points, shown for $\mathrm{Re} = 32$ ($\Box$), 93 ({\Large $\circ$}), and 245 ($\triangle$). The separation $r$ is scaled by the mean magnet spacing $L$. Dashed vertical lines show the positions of some of the peaks expected for a 2D square lattice (other peaks are not observed due to the finite resolution). As  $\mathrm{Re}$ increases, the spatial order vanishes and $g(r)$ becomes liquid-like.}
\end{center}
\end{figure}

To elucidate the transition between the ordered and disordered states of the
topologically special points, we show in Fig.~3 the radial distribution
function $g(r)$ \cite{g_of_r} of the special point positions for three Reynolds
numbers.  $g(r)$ is defined to be the mean number density a distance $r$ from a
fixed position, normalized by the bulk number density. In an ordered state,
$g(r)$ is expected to show a series of peaks corresponding to lattice sites; in
a disordered state, however, $g(r)$ should be unity. At $\mathrm{Re} = 32$,
when the special points are tightly bound to their lattice sites, $g(r)$ is
found to be peaked at many of the locations expected for a 2D square lattice.
As $\mathrm{Re}$ increases, the peaks broaden and $g(r)$ gradually loses
structure. By $\mathrm{Re} = 245$, where the special points move freely and the
spatial order has vanished, $g(r)$ is unity and the special points form a
liquid-like state. To quantify the loss of order, we plot the maximum value of
$g(r)$ (corresponding to the height of its first peak) as a function of
$\mathrm{Re}$ in Fig.~4. 

\begin{figure}
\begin{center}
\includegraphics[width=3.2in]{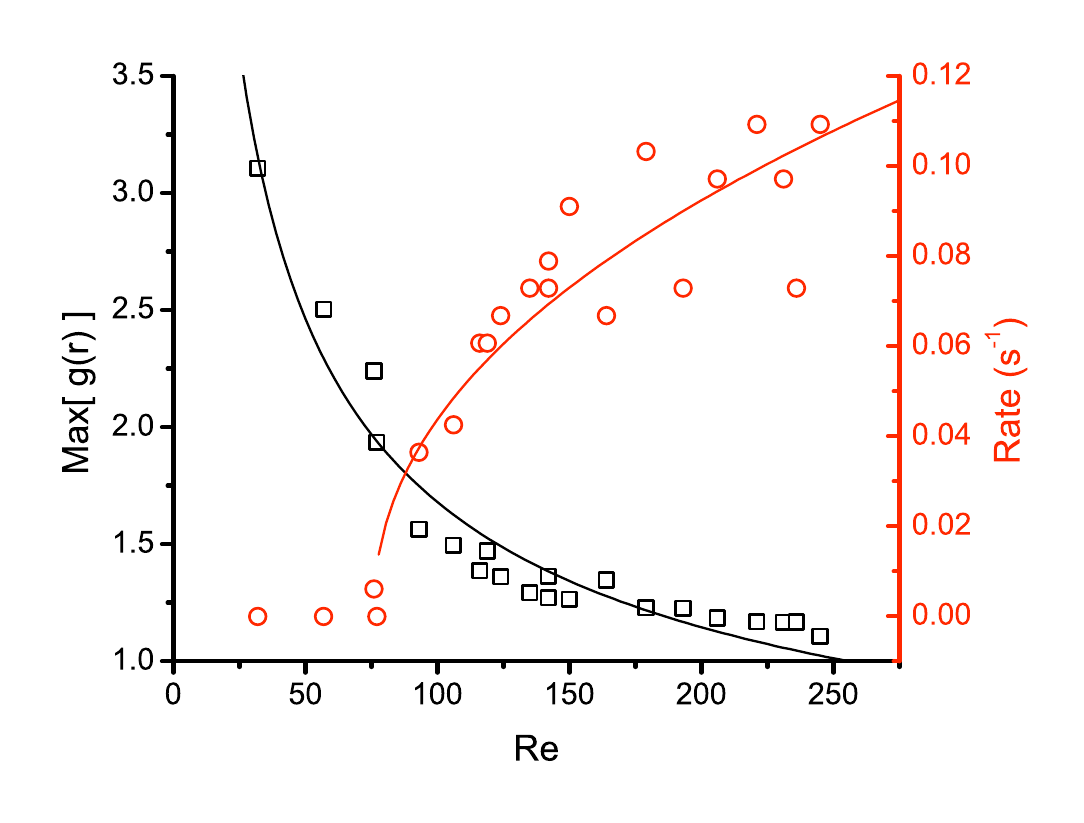}
\caption{\label{fig:rates} (color online) The maximum value of $g(r)$ ($\Box$, left axis), showing a gradual transition to spatial disorder with increasing $\mathrm{Re}$, and the annihilation rate ({\Large $\circ$}, right axis) of hyperbolic-elliptic pairs, which shows a well-defined threshold. The solid lines are drawn to guide the eye.}
\end{center}
\end{figure}

\begin{figure}
\begin{center}
\includegraphics[width=3.2in]{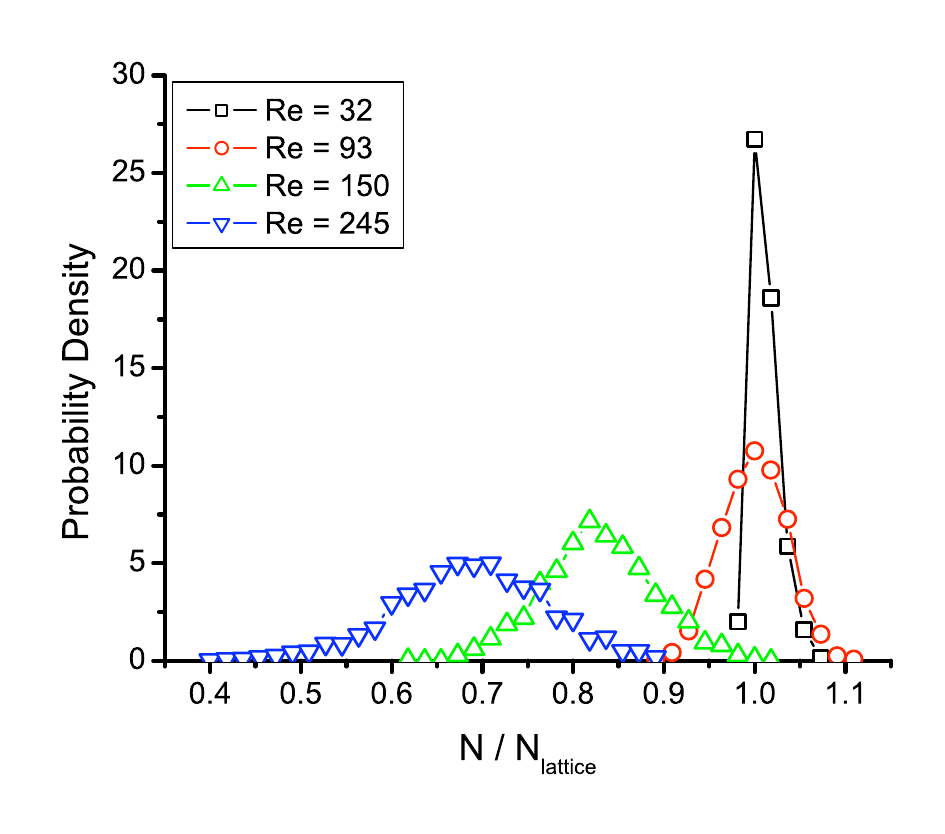}
\caption{\label{fig:numbers} (color online) PDFs of the number $N$ of topologically special points present in the measurement area at each instant in time, normalized by the number of lattice sites, for four Reynolds numbers. As $\mathrm{Re}$ increases, the mean number of special points decreases, while the width of the distribution grows.}
\end{center}
\end{figure}

Once the topologically special points can move appreciably around their lattice
sites, they undergo pairwise interactions that change the topology of the flow.
They can be annihilated in vortex mergers, and new hyperbolic-elliptic pairs
are created when vortices split. By recording the number of such events from
our special point trajectories, we can measure the rates for each of these
processes. The annihilation rate is shown as a function of $\mathrm{Re}$ in
Fig.~4; we find that the creation and annihilation rates are equal to within
experimental accuracy. As expected, these rates grow substantially as the
driving increases.  Surprisingly, however, these rates have a clear onset at a
critical Re, unlike the gradual decline of $g(r)$. We have also measured other
indicators of disorder, including the average of the local speed $\langle
\partial u / \partial t \rangle$ and the mean of the largest Lyapunov exponent
$\langle \lambda \rangle$. As with $g(r)$, however, these measurements show a
continuous change rather than a sharp onset. The lack of a clear threshold in
these measurements may be attributed to the averaging of $g(r)$, $\langle
\partial u / \partial t \rangle$, and $\langle \lambda \rangle$ over space, time, or
both.  In contrast, the creation and annihilation of hyperbolic and elliptic
point pairs are local in both space and time.

The number of special points in the measurement volume changes in tandem with
the rise of the pair creation and annihilation rates. In Fig.~5, we show the
PDFs of the number of special points at each instant in time for several
Reynolds numbers. As $\mathrm{Re}$ increases, the pattern coarsens and the mean
number drops, consistent with the well-known inverse energy cascade in 2D
turbulence \cite{Kraichnan1967}. At the same time, the width of the
distribution grows, which signals the increased activity of the special points
and the corresponding increase in the frequency of topological changes.

In summary, we have developed a method to locate the moving hyperbolic and
elliptic points experimentally by measuring the curvature of particle
trajectories, and have used them to characterize the dynamics of a 2D flow. As
Re is increased, these points gradually unbind from their preferred locations
(determined by the forcing), deforming the entire flow pattern. When they are
created or annihilated in pairs, starting at $Re \sim 75$, they change the
topology of the pattern. The behavior of these points shows that the transition
to STC involves two successive stages: a gradual loss of spatial and temporal
order followed by a surprising abrupt onset of topological changes. We suggest
that new theories of STC may be developed using these topological ideas.

\begin{acknowledgments}
We thank D. Egolf, P. Love, H. Riecke, and G. Voth for helpful discussions. This research was supported by the U.S.~National Science Foundation under Grant DMR-0405187. 
\end{acknowledgments}

\bibliography{STC}

\end{document}